\documentclass{vgtc}

\ifpdf
  \pdfoutput=1\relax                   
  \usepackage{graphicx}                
  \DeclareGraphicsExtensions{.pdf,.png,.jpg,.jpeg} 
\else                                 
  \usepackage{graphicx}                
  \DeclareGraphicsExtensions{.eps}     
\fi

\graphicspath{{figures/}{pictures/}{images/}{./}}

\usepackage{microtype}
\PassOptionsToPackage{warn}{textcomp}
\usepackage{textcomp}
\usepackage{mathptmx}
\usepackage{times}

\usepackage{cite}
\usepackage{tabu}
\usepackage{booktabs}
\usepackage[dvipsnames]{xcolor}
\usepackage{xspace}
\usepackage{comment}
\usepackage{enumitem}
\usepackage{tabu}
\usepackage{rotating}
\usepackage{tikz}
\usepackage{capt-of,etoolbox}
\usepackage{microtype}
\usepackage{bm}
\usepackage{soul}
\usepackage{wrapfig}
\usepackage{graphbox}
\usepackage{tcolorbox}
\usepackage{amssymb}
\usepackage{amsmath}
\usepackage{mathtools}
\usepackage{balance}
\usepackage[varqu]{zi4}
\usepackage[bb=boondox]{mathalfa}
\usepackage[utf8x]{inputenc}
\usepackage{units}
\usepackage{setspace}
\usepackage{gensymb}
\usepackage{helvet}
\usepackage{printlen}

\newcommand{\tool}{\textsc{Visual Auditor}}
\definecolor{blueVI}{RGB}{30, 136, 229}
\newcommand{\link}[1]{{\href{#1}{\color{blueVI}\textbf{\texttt{#1}}}}}

\onlineid{0}

\vgtccategory{Research}
\vgtcpapertype{application/design study}

\vgtcinsertpkg

\title{\tool{}: Interactive Visualization for Detection and Summarization of Model Biases}

\renewcommand\footnotemark{}
\newcommand{\authorgap}{\hspace{10pt}}

\author{
  David Munechika\textsuperscript{\textrm 1}
  \thanks{\textsuperscript{\textrm 1}Georgia Tech. \{\href{mailto:david.munechika@gatech.edu}{dmunechika3}$\mid$\href{jayw@gatech.edu}{jayw}$\mid$\href{mailto:polo@gatech.edu}{polo}\}@gatech.edu} \authorgap
  Zijie J. Wang\textsuperscript{\textrm 1} \authorgap
  Jack Reidy\textsuperscript{\textrm 2}
  \thanks{\textsuperscript{\textrm 2}Fiddler AI. \{\href{mailto:jack@fiddler.ai}{jack}$\mid$\href{mailto:josh@fiddler.ai}{josh}$\mid$\href{mailto:krishna@fiddler.ai}{krishna}$\mid$\href{mailto:krishnaram@fiddler.ai}{krishnaram}\}@fiddler.ai} \authorgap
  Josh Rubin\textsuperscript{\textrm 2} \authorgap
  Krishna Gade\textsuperscript{\textrm 2} \authorgap \\
  Krishnaram Kenthapadi\textsuperscript{\textrm 2} \authorgap
  Duen Horng Chau\textsuperscript{\textrm 1} \authorgap
}

\teaser{
  \centering
  \makebox[\textwidth][c]{\includegraphics[width=.95\linewidth]{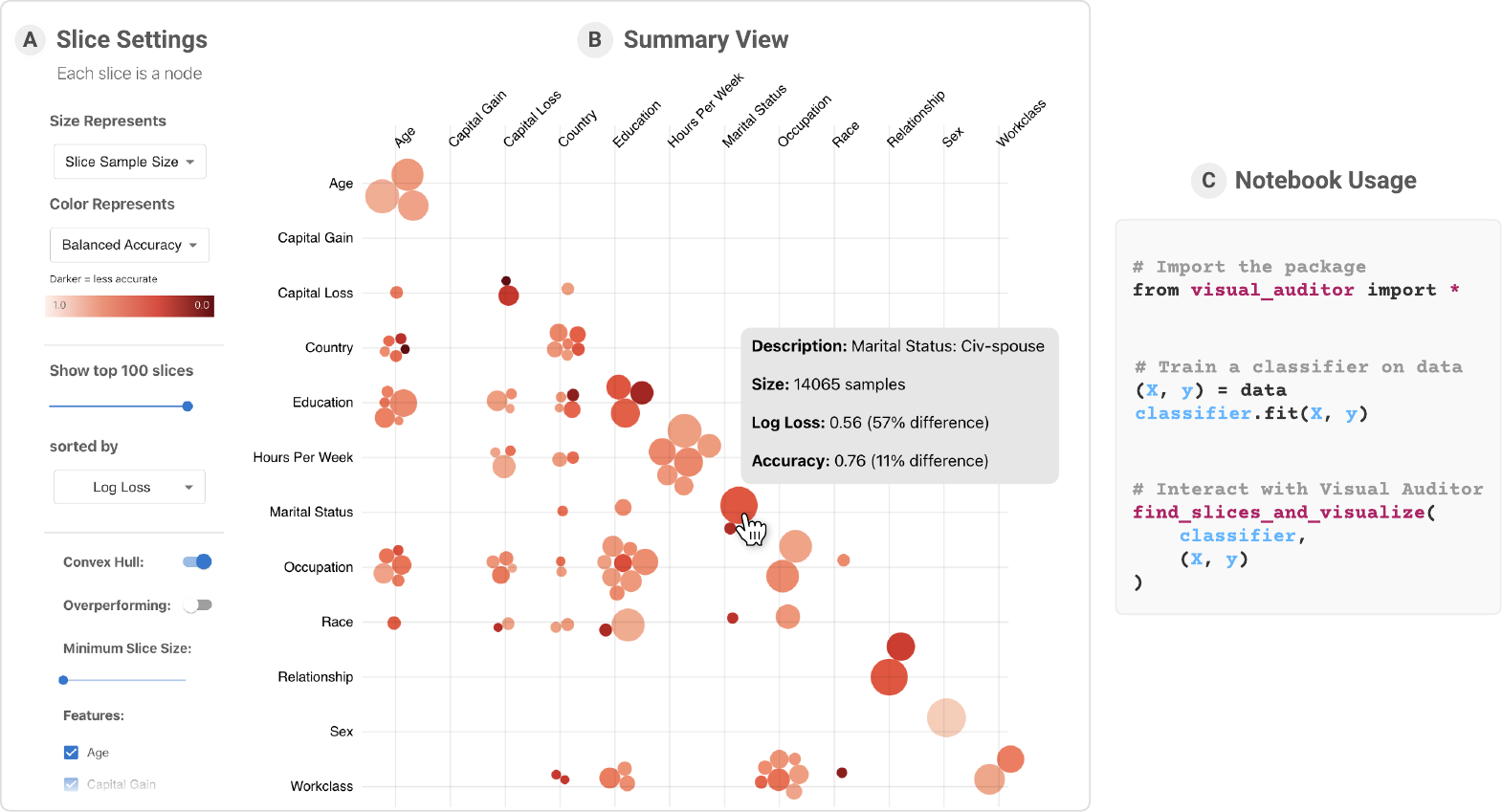}}
    \caption{\tool{} provides an overview of underperforming data slices to show where intersectional biases exist. 
    \textbf{(A)} The \textit{Slice Settings} sidebar contains options for filtering the data and modifying the visualization, such as customizing the size and color encodings, showing the top \textit{k} slices, or selecting particular features of interest. 
    \textbf{(B)} The \textit{Summary View} provides a visual overview of the underperforming data slices. Here currently displays the \textit{Force Layout} which shows underperforming data slices as nodes on a grid. The location of each node is determined by the features that define the data slice. Users can view clusters of similar data slices to better understand where intersectional bias might exist in their model. \textbf{(C)} \tool{} is an open-source tool that easily integrates within existing data science workflows and can be accessed directly within computational notebooks.}
  \label{fig:teaser}
}

\abstract{As machine learning (ML) systems become increasingly widespread, it is necessary to audit these systems for biases prior to their deployment. Recent research has developed algorithms for effectively identifying intersectional bias in the form of interpretable, underperforming subsets (or slices) of the data. However, these solutions and their insights are limited without a tool for visually understanding and interacting with the results of these algorithms. We propose \tool{}, an interactive visualization tool for auditing and summarizing model biases. \tool{} assists model validation by providing an interpretable overview of intersectional bias (bias that is present when examining populations defined by multiple features), details about relationships between problematic data slices, and a comparison between underperforming and overperforming data slices in a model. Our open-source tool runs directly in both computational notebooks and web browsers, making model auditing accessible and easily integrated into current ML development workflows. An observational user study in collaboration with domain experts at Fiddler AI highlights that our tool can help ML practitioners identify and understand model biases.}

\CCScatlist{
  \CCScatTwelve{Human-centered computing}{Visualization}{}{}
}

\begin{document}

\maketitle

\section{Introduction}

The growing success and popularity of machine learning (ML) has led to widespread applications in the real world. It is therefore necessary to ensure that deployed systems exhibit fair treatment across all subgroups of people \cite{barocas2016big}. Without proper model auditing and validation, we risk encoding prejudicial biases into our models, thereby deploying systems in the real world that reflect our social biases and result in user discrimination \cite{mehrabi2021survey, barocas2016big}. 

An example can be found at the forefront of the ML fairness debate, specifically surrounding the algorithms behind recidivism prediction instruments (RPIs). Studies have found the predictive accuracy of certain RPIs, such as the COMPAS \cite{northpoint2012} or PCRA \cite{pcra2018} instruments, vary significantly between different demographic groups, serving as evidence of algorithmic bias \cite{chouldechova2016fair, skeem2016risk, Flores2016FalsePF}. Similar concerns about perpetuating discrimination through algorithmic unfairness have been raised in other areas including credit card scoring, housing advertisements, and mortgage systems \cite{allen2019color}.
Often, a high overall model performance can cover up the unsatisfactory performance of individual subgroups (or slices) of data \cite{sagadeeva2021, polyzotis2019, cabrera2019}. 

To mitigate biases, recent researchers often identify slices of data that appear particularly problematic—these are characterized by the ML model exhibiting unusually poor performance on the slice compared to the model's overall performance and the slice being large enough to be significant rather than simply an outlier \cite{polyzotis2019, sagadeeva2021}. While existing algorithms \cite{polyzotis2019, sagadeeva2021} are capable of identifying problematic slices, they fail to provide effective \cite{saha2020measuring}, interpretable \cite{lee2021landscape} overviews of the model bias. Furthermore, existing tools \cite{wexler2020googleWhatIf, bird2020fairlearn, bellamy2018aiFairness360} that aim to illustrate model performance and examine discrimination in ML models require users to have prior knowledge about which biases exist within the model.

To address these limitations, we design and develop \tool{}, an interactive, visual interface for model bias detection and summarization. We aim to create a visual overview that enables ML models to be audited so that underperforming subgroups can be effectively surfaced and analyzed by human users. Our research makes the following major contributions:

\begin{itemize}[topsep=1pt, itemsep=0mm, parsep=3pt, leftmargin=9pt]
  \item 
  \itemsep0em 
  \textbf{\tool{}, an interactive visualization tool} for auditing and summarizing ML model biases. \tool{} fills a critical research gap and practical need for visually summarizing and auditing underperforming data slices. Underperforming subgroups are identified using slice-finding algorithms \cite{polyzotis2019} and are visually displayed along with important accompanying information including the size of the subgroup, how significantly it is underperforming, and what features it is defined by. \tool{} also complements existing research that focuses primarily on algorithmic slice finding \cite{polyzotis2019, sagadeeva2021}. It goes beyond simply providing details about each data slice by identifying relationships between similar problematic data slices and providing methods for filtering these results to match the interest of the user.
  
  \item \textbf{Design lessons distilled from a user study} with Fiddler AI engineers and data scientists. An observational user study with 4 domain experts highlights how \tool{} may be useful when integrated within existing ML and data science workflows. We discuss design lessons from our iterative design process and the user study results.
  
  \item \textbf{An open-source\footnote{Code: \link{https://github.com/poloclub/visual-auditor}} and web-based implementation} to empower ML practitioners to audit their models for bias. We developed \tool{} with modern web technologies so that anyone can access our tool directly in a web browser or computational notebook. A demo video of \tool{} can be viewed at \link{https://youtu.be/ZGCVtu2fcbc}.
\end{itemize}

\noindent It is our goal that \tool{} will encourage and improve the process of model validation while better enabling the analysis and eventual mitigation of intersectional bias in models.

\section{Background \& Related Work}
\textbf{Slice-Finding Algorithms.}
Auditing models trained on data defined by numerous features is a non-trivial task. One must not only look at subgroups defined by each particular feature but also consider the intersectional bias which may result when defining subgroups by multiple features. The number of potential subgroups of data grows combinatorially, rendering it impractical for a human to perform this type of model validation manually \cite{sagadeeva2021}.

Existing research aims to address these computational issues of data slicing for model validation. SliceFinder is one existing framework for identifying problematic data slices. It uses statistical techniques to find interpretable slices as opposed to arbitrary subsets which are commonly the result of traditional clustering techniques. SliceFinder considers how significant the difference in loss is between the slice and the model itself as well as how large the slice is. It computes an effect size for each data slice which determines how significant (in other words, problematic) a particular data slice is for model validation—a higher effect size indicates higher significance. This effect size metric is used to rank data slices and determine which are the most problematic \cite{polyzotis2019}.

Similarly, SliceLine is an enumeration algorithm which attempts to extend the SliceFinder algorithm by addressing the scalability limitations of traditional methods. It leverages efficient sparse linear algebra to enable slice enumeration even in complex datasets \cite{sagadeeva2021}.

\textbf{Visualization for Model Performance and Bias.}
The objectives of ML fairness and understanding model performance has inspired a wealth of literature \cite{hohman2018visual, patel2008investigating}. Proposed systems have emerged with the intention of elucidating the behavior and interior of models and improving the fairness of AI systems using various visualization techniques. Uber's Manifold \cite{zhang2018manifold} and ModelTracker \cite{amershi2015modeltracker} are two systems that have been used in practice to provide performance insights and comparisons for ML models. However, these tools are solely focused on performance analysis without a specific aim on identifying biases or mitigating algorithmic unfairness.

Other recent work has emerged within the visualization community to specifically address model fairness. Audit AI \cite{auditAI}, Google's What-If tool \cite{wexler2020googleWhatIf}, Microsoft's Fairlearn project \cite{bird2020fairlearn}, and IBM's AI Fairness 360 toolkit \cite{bellamy2018aiFairness360} are all different solutions aiming to mitigate fairness issues within AI and ML models. They provide details about fairness metrics of different subgroups of data and use various statistical methods for comparing different groups. However, these systems are limited by requiring a priori knowledge of the discriminated groups. Users need to manually identify the protected attributes as well as the privileged and unprivileged groups in order for these systems to generate comparisons and evaluate potential biases. It is infeasible for users to manually consider all potential subgroups of data in order to identify biases \cite{sagadeeva2021}—therefore, without combining bias detection with these visualization methods, these tools are not easily integrated into current ML workflows.

The SliceFinder GUI attempts to solve this issue by combining a user interface with the SliceFinder algorithm \cite{chung2019slice}. However, this visual system only presents the textual outputs of the algorithm in a table without offering additional details about the data slices. It lacks capabilities to analyze problematic slices or understand model bias at a higher level.
In contrast, our tool synthesizes bias detection with interpretable visualizations to address the limitations of these existing systems. It effectively summarizes data slices, allows filtering by desirable characteristics, displays fairness metrics, illustrates related overlapping slices as well as clusters of similarly-defined slices, and is easily integrated within current workflow environments.

\section{Design Goals}
Through close collaboration with engineers and scientists at Fiddler AI since August 2021, we have learned about the need for 
an interactive visual tool 
that helps ML practitioners 
summarize and analyze model bias. 
Through a literature review, we have identified four design goals of \tool{}.

\begin{enumerate}[topsep=2pt, itemsep=0mm, parsep=3pt, leftmargin=19pt, label=\textbf{G\arabic*.}, ref=G\arabic*]
  \item \label{item:summary}
  \textbf{Visual summary of problematic data slices.}
  Depending on the hyperparameters of the slice-finding algorithm (such as the degree, effect size threshold, or maximum number of slices), the number of problematic slices found by slice-finding algorithms can vary from only a few to over a hundred\cite{polyzotis2019}. Without effective visual techniques, it is challenging for users to understand and explore existing bias in their model \cite{rudinInterpretableMachineLearning2022}. Therefore, we designed scalable visualizations to summarize the problematic data slices to help users better understand where their model is underperforming ~(\autoref{sec:interface:force}).

  \item \label{item:filter}
  \textbf{Ability to filter slices by desirable characteristics.}
  To effectively audit their models, users need the ability to narrow their focus down to problematic slices that are of particular interest to them \cite{rudinInterpretableMachineLearning2022}. These might be slices with important and potentially sensitive features \cite{dongExploringCloudVariable2020}, slices of a particular size (to eliminate outliers), or slices characterized by an unusually high or low associated fairness metric \cite{dblpImpossibilityTheorem2020}~(\autoref{sec:interface:filters}).

  \item \label{item:comparing}
  \textbf{Comparing slices and analyzing slice relationships.}
  Examining similar data slices can be useful to understand feature importance or identify larger, generalized slices (as the result of merging similar slices) \cite{cabrera2019, dwork2011fairnessAwareness, zemel2013learningFair}. This would be useful for identifying efficient bias mitigation techniques that target clusters of similar problematic slices in order to yield the greatest improvement in model performance
  ~(\autoref{sec:interface:graph}). The inclusion of overperforming slices for comparison can also be useful, especially in inbalanced datasets \cite{chawla2009imbalancedDatasets}. Comparing the differences in characteristics between underperforming and overperforming slices can yield useful insights regarding the performance of a particular model in order to identify strategies for mitigating existing model bias  \cite{he2009imbalancedData, kubat1997curseImbalanced, batista2004methodsBalancing}~(\autoref{sec:interface:overperforming}).

  \item \label{item:implementation}
  \textbf{Integrating into common development workflows.}
  Modern data science workflows rely on the use of computational notebooks such as Jupyter Notebook \cite{kluyverJupyterNotebooksPublishing2016}. Data scientists and ML practitioners frequently work within computational notebooks to develop and train ML models \cite{perkelReactiveReproducibleCollaborative2021}. To ensure model auditing is accessible within the current workflow, we designed \tool{} to run in a web browser and as an interactive widget in computational notebooks.
  Finally, we open-sourced our tool to encourage and support future design, research, and development of model auditing and bias mitigation~(\autoref{sec:interface:implementation}).

\end{enumerate}

\section{System Design}
Following the design goals, \tool{} (Fig. 1) tightly integrates
four components: the \textit{Force Layout}~(\autoref{sec:interface:force}), the \textit{Slice Settings} panel~(\autoref{sec:interface:filters}), the \textit{Graph Layout}~(\autoref{sec:interface:graph}), and the \textit{Overperforming Slices}~(\autoref{sec:interface:overperforming}).

\subsection{Summarizing the Problematic Slices}
\label{sec:interface:force}
\tool{} uses the SliceFinder \cite{chung2019slice} algorithm in the backend for identifying problematic data slices but extends its interpretability and actionability. We limit intersectional biases to be defined by at most two features to ensure interpretable slices with a significant effect size are identified. To help users efficiently understand and analyze the potentially large number of problematic slices produced by the algorithm, we present these in a summary view.

\begin{figure}[tb]
  \centering
  \makebox[\linewidth][c]{\includegraphics[width=1.1\linewidth]{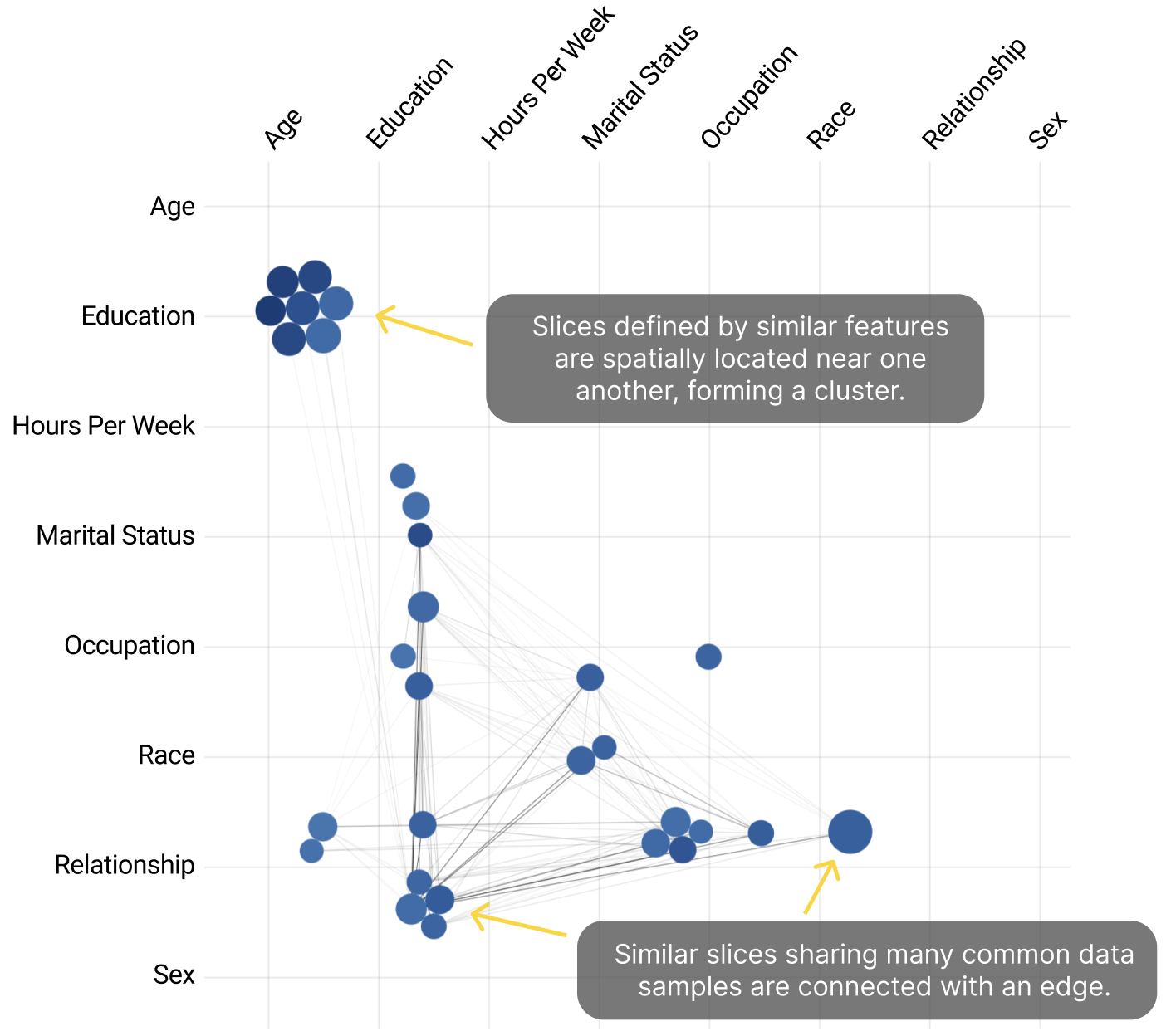}}
  \caption{Users can also view overperforming slices to compare and learn from. This figure shows the \textit{Graph Layout} which connects two slices with an edge if they share many data instances. Thicker edges indicate stronger relationships between slices and a darker node color indicates higher performance compared to the overall model.}
  \label{fig:overperforming}
\end{figure}

\textbf{Force Layout.} The \textit{Force Layout} (Figure \ref{fig:teaser}B)  summarizes the problematic slices in a dataset~(\ref{item:summary}). Each slice is displayed as a node on a grid and mapped to an area defined by the intersection of its features. Slices will be spatially located next to other slices defined by similar features. By default, the color of each node maps to the percent difference of its log loss compared to the overall model (a darker color indicates worse log loss and more severe underperformance). Similarly, the size of each node represents the sample size of the slice standardized on a log scale. However, both the color and size encodings can be customized based on user preference. Users can also hover over a specific node to display a tooltip containing details about that particular slice. Overall, the \textit{Force Layout} is an effective visualization design because it immediately draws attention to the largest and most problematic slices through color and size encodings (\ref{item:summary}) while simultaneously conveying information about relationships through clusters of similarly defined slices (\ref{item:comparing}).

\subsection{Slice Filters and Visualization Settings}
\label{sec:interface:filters}
The \textit{Force Layout} by default displays an overview of all of the slices. To enable users to focus on particular slices of interest~(\ref{item:filter}), the \textit{Slice Settings} sidebar (Figure \ref{fig:teaser}A) offers a collection of options for modifying the visualization and filtering the visible slices. 

The \textit{Color Represents} and \textit{Size Represents} dropdown menus allow the user to customize each of these encodings. For example, the respective default encodings of log loss and sample size could be changed to a fairness metric such as balanced accuracy, if desired. Additionally, the \textit{Show Top \textsf{k} Slices} slider and \textit{Sorted By} dropdown allow users to specify a dimension (e.g. sample size, log loss, balanced accuracy) to sort the slices by and filter down to only the top \textit{\textsf{k}} slices for that particular dimension. 

The \textit{Minimum Slice Size} slider sets a threshold for the smallest allowed slice sample size which is useful to filter out outlier slices that are composed of only a few samples and not necessarily representative of an existing bias. Similarly, the \textit{Features} checkboxes enable users to specify features of interest and only display slices that include these features in their definitions. This is particularly helpful when dealing with model fairness because usually some attributes will be considered ``protected'' or more sensitive than others \cite{zhe2021protected}.

These various filtering options allow users to quickly identify the most problematic slices within their data and slices that are of particular importance to them. This effectively speeds up the process of identifying the most significant slices hurting model performance and understanding how to mitigate the existing bias.

\subsection{Similar Slice Relationships}
\label{sec:interface:graph}
\textbf{Graph Layout.} While the \textit{Force Layout} displays slices with similar features in distinct clusters on a grid, it does not allow the user to identify similar slices and relationships between slices. We define similar slices to be those that share a high number of data instances (i.e. overlapping samples) \cite{cabrera2019}. To display the relationships between slices~(\ref{item:comparing}), we design the \textit{Graph Layout} (Figure \ref{fig:overperforming}) which is an extension of the \textit{Force Layout} that includes edges connecting nodes which represent similar data slices.

The number of overlapping samples between two slices determines the thickness of that particular edge, and the strength of the force pulling the nodes together. This results in nodes that share many common data samples being clustered closer together. The \textit{Edge Force Strength} slider can be used to control the strength of the force of attraction between two connected nodes (with an edge force strength of zero degenerating into the \textit{Force Layout}). Furthermore, the \textit{Edge Filtering} slider can set the minimum number of overlapping samples necessary for an edge to exist between two nodes.

\subsection{Overperforming Slices}
\label{sec:interface:overperforming}
Viewing which intersections of features yield the highest accuracies presents users with an additional insight into the performance of their model. To support the analysis of overperforming data slices~(\ref{item:comparing}), we automatically compute these slices as part of the algorithm when the problematic slices are found. Users can switch to viewing overperforming slices by toggling on the \textit{Overperforming} switch in the \textit{Slice Settings} sidebar when viewing any visualization (Figure \ref{fig:overperforming}).

\subsection{Accessible, Open-source Implementation}
\label{sec:interface:implementation}
\tool{} is a web-based, interactive visualization tool built with D3.js \cite{bostock2011} and React.js \cite{react}. Users can access the tool using a web browser or directly within a computational notebook (\ref{item:implementation}). To increase the accessibility of our tool, we have released \tool{} on the Python Package Index (PyPI) \footnote{PyPI Package: \href{https://pypi.org/project/visual\_auditor/}{\textcolor{blueVI}{\textbf{https://pypi.org/project/visual\_auditor/}}}}. Computing slices and generating the visual interface can be done in only a few lines of code (Figure \ref{fig:teaser}C). We also open-sourced our implementation so future researchers can extend the tool's design and functionality.

\section{User Study}
\begin{figure}[tb]
  \centering
  \includegraphics[width=1\linewidth]{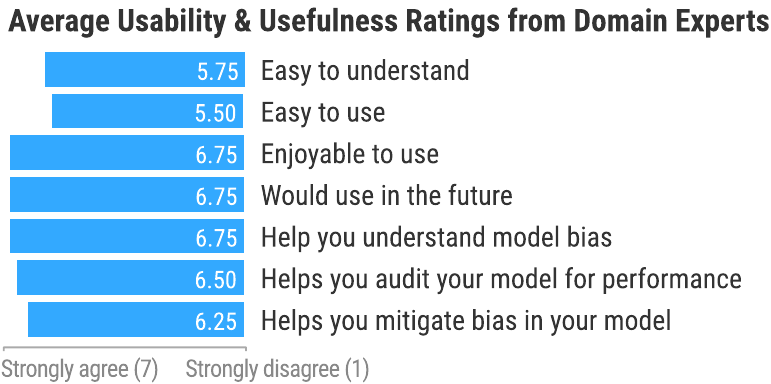}
  \caption{In our user evaluation, participants thought \tool{} was easy and enjoyable to use. The domain experts rated the usefulness of the tool very favorably, providing support that our interactive tool is effective at summarizing and analyzing model bias.}
  \label{fig:ratings}
\end{figure}
The \tool{} project started in August 2021. Internal design iterations with two data scientists from Fiddler AI were performed to gather feedback and further refine the tool. Finally, a user study was conducted with 4 additional data scientists and engineers (P1-P4) to investigate the effectiveness and usability of \tool{}. All participants were domain experts who indicated the highest level of familiarity with ML and data science. Participants received no compensation for taking part in the study.

\smallskip
\noindent
\textbf{Data.} 
Participants were given the option of working with either the Adult Census Income dataset \cite{kohavi1996adult}, which predicts whether or not an individual's income exceeds \$50k per year, or the Statlog (German Credit) dataset \cite{dheeru2017germanCredit}, which classifies people as good or bad credit risks according to a set of attributes. Both of these datasets are popularly used in research related to algorithmic fairness \cite{bellamy2018aiFairness360, verma2018fairness, kearns2019empirical}.

\smallskip
\noindent
\textbf{Procedure.}
The study duration was one hour per participant. In each study, participants was invited to an individual Zoom meeting. First, they were asked to fill out a pre-study background survey and sign a consent form. They were given a brief tutorial that explained the purpose of \tool{} and demonstrated its capabilities. Following the tutorial, the participants selected a dataset to use and were provided with the relevant feature names and textual descriptions. They were also given the link to a Jupyter Notebook hosted on Binder\footnote{User Study Binder: \href{https://mybinder.org/v2/gh/davidmunechika/visual-auditor-demo-repo/master}{\textcolor{blueVI}{\textbf{https://mybinder.org/v2/gh/davidmunechika/visual-auditor-demo-repo/master}}}} for testing \tool{} with their particular dataset. Within this simulated data science workflow, participants were asked to explore \tool{} and analyze problematic slices in their model while thinking aloud. The exploration of the tool was screen-recorded over Zoom. During this meeting they were informed that they could ask design- or functionality-related questions to the researchers. Each session ended with a usability questionnaire and post-study exit survey to finalize their evaluation.

\smallskip
\noindent
\textbf{Results.}
Average ratings for \tool{}'s usability and usefulness are shown in Figure \ref{fig:ratings}. All participants found the tool to be easy to understand and enjoyable to use. The domain experts also agreed that this tool provides new functionality for understanding model bias, auditing models for performance, and finding approaches to mitigate ML bias that did not previously exist.

Through the user study, we learned that the \textbf{\textit{Force Layout} was an effective visual design for summarizing model bias}. P2 commented, \textit{``the key feature I found most helpful was being able to visualize sample size and performance of all the slices at once. My eye was immediately drawn to underperforming slices with a large sample size.''} This layout was also interpretable and understandable for participants as they found the visual representation of slices to be inherently intuitive. P3 noted that \tool{} provided a \textit{``great visual representation of problematic slices [making] it easier to debug the model. The tool did a great job of representing sizes and relative performances of the slices using colors, so I was able to quickly understand and use the tool.''}

\textbf{Participants appreciated the sample size scaling and slice filtering options.} Multiple participants identified the ability to filter the top \textit{\textsf{k}} slices and set a minimum slice size as one of the most useful features. P1 said these filters helped them \textit{``hone in on the most problematic slices''}. Other participants found the \textit{Graph Layout} to be effective at identifying relationships between slices. P1 was particularly interested in examining the groups of densely connected nodes in the \textit{``Graph Layout to target a bigger subsection.''} Their strategy was to identify generalized data slices connecting multiple problematic slices so that targeting these slices would result in the greatest performance improvement in the model.

Lastly, the \textbf{notebook support and ease of integration into current workflows was viewed very favorably by participants}. From an initial background survey, all participants rated their notebook usage at either the highest level (``every day'') or next highest level (``very frequently''). Every participant expressed an interest in using \tool{} in the future, and one participant (P4) with expertise specifically in mitigating intersectional fairness issues said they would \textit{``absolutely use this tool in [their] visualization work.''}

\section{Conclusion \& Discussion}
Mitigating bias and maximizing performance should be important considerations during the development of ML models. \tool{} addresses these issues by providing an overview of problematic data slices through the use of interactive visualization techniques. Our user study has shown that \tool{} is effective at summarizing and analyzing model bias and can be easily integrated into existing development workflows. For future work, we plan to extend \tool{}'s capabilities to offer more actionable methods for mitigating existing bias. Intersectional bias can exist in a model for various reasons, including insufficient training data, explicitly or implicitly encoded social biases, or an overly-simple model architecture. A tool that identifies which solution would be most effective for addressing model bias would make the model validation process more efficient. We also plan to enhance \tool{} by improving graph view readability and allowing for the summarization of intersectional bias generated from more than two features. By interactively exploring the intersectional bias that exists in a model and providing effective bias mitigation strategies, \tool{} will help to ensure models deployed in the real world exhibit fair treatment across all subgroups of people.

\clearpage
\bibliographystyle{abbrv-doi}

\bibliography{references}

\begin{thebibliography}{10}

\bibitem{auditAI}
Audit ai.
\newblock \url{https://github.com/pymetrics/audit-ai}.
\newblock Accessed: 2022-04-25.

\bibitem{react}
React.js.
\newblock \url{https://github.com/facebook/react}.
\newblock Accessed: 2022-04-25.

\bibitem{allen2019color}
J.~A. Allen.
\newblock The color of algorithms: An analysis and proposed research agenda for
  deterring algorithmic redlining.
\newblock {\em Fordham Urb. LJ}, 46:219, 2019.

\bibitem{amershi2015modeltracker}
S.~Amershi, M.~Chickering, S.~M. Drucker, B.~Lee, P.~Simard, and J.~Suh.
\newblock Modeltracker: Redesigning performance analysis tools for machine
  learning.
\newblock In {\em Proceedings of the 33rd Annual ACM Conference on Human
  Factors in Computing Systems}, pp. 337--346, 2015.

\bibitem{barocas2016big}
S.~Barocas and A.~D. Selbst.
\newblock Big data's disparate impact.
\newblock {\em Calif. L. Rev.}, 104:671, 2016.

\bibitem{batista2004methodsBalancing}
G.~E. A. P.~A. Batista, R.~C. Prati, and M.~C. Monard.
\newblock A study of the behavior of several methods for balancing machine
  learning training data.
\newblock {\em SIGKDD Explor. Newsl.}, 6(1):20–29, jun 2004. doi: {{%
10\hspace{.1pt}\discretionary{.}{%
}{.}\hspace{.4pt}1145\discretionary{/}{%
}{/}1007730\hspace{.1pt}\discretionary{.}{%
}{.}\hspace{.4pt}1007735}}


\bibitem{bellamy2018aiFairness360}
R.~K. Bellamy, K.~Dey, M.~Hind, S.~C. Hoffman, S.~Houde, K.~Kannan, P.~Lohia,
  J.~Martino, S.~Mehta, A.~Mojsilovic, et~al.
\newblock Ai fairness 360: An extensible toolkit for detecting, understanding,
  and mitigating unwanted algorithmic bias.
\newblock {\em arXiv preprint arXiv:1810.01943}, 2018.

\bibitem{bird2020fairlearn}
S.~Bird, M.~Dud{\'\i}k, R.~Edgar, B.~Horn, R.~Lutz, V.~Milan, M.~Sameki,
  H.~Wallach, and K.~Walker.
\newblock Fairlearn: A toolkit for assessing and improving fairness in ai.
\newblock {\em Microsoft, Tech. Rep. MSR-TR-2020-32}, 2020.

\bibitem{bostock2011}
M.~Bostock, V.~Ogievetsky, and J.~Heer.
\newblock D³ data-driven documents.
\newblock {\em IEEE Transactions on Visualization and Computer Graphics},
  17(12):2301--2309, 2011. doi: {{%
10\hspace{.1pt}\discretionary{.}{%
}{.}\hspace{.4pt}1109\discretionary{/}{%
}{/}TVCG\hspace{.1pt}\discretionary{.}{%
}{.}\hspace{.4pt}2011\hspace{.1pt}\discretionary{.}{%
}{.}\hspace{.4pt}185}}


\bibitem{cabrera2019}
{\'{A}}.~A. Cabrera, W.~Epperson, F.~Hohman, M.~Kahng, J.~Morgenstern, and
  D.~H. Chau.
\newblock Fairvis: Visual analytics for discovering intersectional bias in
  machine learning.
\newblock {\em CoRR}, abs/1904.05419, 2019.

\bibitem{chawla2009imbalancedDatasets}
N.~V. Chawla.
\newblock Data mining for imbalanced datasets: An overview.
\newblock {\em Data mining and knowledge discovery handbook}, pp. 875--886,
  2009.

\bibitem{chouldechova2016fair}
A.~Chouldechova.
\newblock Fair prediction with disparate impact: A study of bias in recidivism
  prediction instruments, 2016.

\bibitem{chung2019slice}
Y.~Chung, T.~Kraska, N.~Polyzotis, K.~H. Tae, and S.~E. Whang.
\newblock Slice finder: Automated data slicing for model validation.
\newblock In {\em 2019 IEEE 35th International Conference on Data Engineering
  (ICDE)}, pp. 1550--1553. IEEE, 2019.

\bibitem{dheeru2017germanCredit}
D.~Dheeru and E.~Karra~Taniskidou.
\newblock Uci machine learning repository.
\newblock
  \url{https://archive.ics.uci.edu/ml/datasets/Statlog+\%28German+Credit+Data\%29},
  2017.
\newblock Accessed: 2022-04-25.

\bibitem{dongExploringCloudVariable2020}
J.~Dong and C.~Rudin.
\newblock Exploring the cloud of variable importance for the set of all good
  models.
\newblock {\em Nature Machine Intelligence}, 2(12), 2020.

\bibitem{dwork2011fairnessAwareness}
C.~Dwork, M.~Hardt, T.~Pitassi, O.~Reingold, and R.~Zemel.
\newblock Fairness through awareness, 2011. doi: {{%
10\hspace{.1pt}\discretionary{.}{%
}{.}\hspace{.4pt}48550\discretionary{/}{%
}{/}ARXIV\hspace{.1pt}\discretionary{.}{%
}{.}\hspace{.4pt}1104\hspace{.1pt}\discretionary{.}{%
}{.}\hspace{.4pt}3913}}


\bibitem{Flores2016FalsePF}
A.~W. Flores, K.~A. Bechtel, and C.~T. Lowenkamp.
\newblock False positives, false negatives, and false analyses: A rejoinder to
  "machine bias: There's software used across the country to predict future
  criminals. and it's biased against blacks".
\newblock {\em Federal Probation}, 80:38, 2016.

\bibitem{he2009imbalancedData}
H.~He and E.~A. Garcia.
\newblock Learning from imbalanced data.
\newblock {\em IEEE Transactions on Knowledge and Data Engineering},
  21(9):1263--1284, 2009. doi: {{%
10\hspace{.1pt}\discretionary{.}{%
}{.}\hspace{.4pt}1109\discretionary{/}{%
}{/}TKDE\hspace{.1pt}\discretionary{.}{%
}{.}\hspace{.4pt}2008\hspace{.1pt}\discretionary{.}{%
}{.}\hspace{.4pt}239}}


\bibitem{hohman2018visual}
F.~Hohman, M.~Kahng, R.~Pienta, and D.~H. Chau.
\newblock Visual analytics in deep learning: An interrogative survey for the
  next frontiers.
\newblock {\em IEEE transactions on visualization and computer graphics},
  25(8):2674--2693, 2018.

\bibitem{kearns2019empirical}
M.~Kearns, S.~Neel, A.~Roth, and Z.~S. Wu.
\newblock An empirical study of rich subgroup fairness for machine learning.
\newblock In {\em Proceedings of the conference on fairness, accountability,
  and transparency}, pp. 100--109, 2019.

\bibitem{kluyverJupyterNotebooksPublishing2016}
T.~Kluyver and {others}.
\newblock Jupyter {{Notebooks}} - a publishing format for reproducible
  computational workflows.
\newblock {\em ELPUB}, 2016.

\bibitem{kohavi1996adult}
R.~Kohavi et~al.
\newblock Scaling up the accuracy of naive-bayes classifiers: A decision-tree
  hybrid.
\newblock In {\em Kdd}, vol.~96, pp. 202--207, 1996.

\bibitem{kubat1997curseImbalanced}
M.~Kubat, S.~Matwin, et~al.
\newblock Addressing the curse of imbalanced training sets: one-sided
  selection.
\newblock In {\em Icml}, vol.~97, p. 179. Citeseer, 1997.

\bibitem{lee2021landscape}
M.~S.~A. Lee and J.~Singh.
\newblock The landscape and gaps in open source fairness toolkits.
\newblock In {\em Proceedings of the 2021 CHI conference on human factors in
  computing systems}, pp. 1--13, 2021.

\bibitem{mehrabi2021survey}
N.~Mehrabi, F.~Morstatter, N.~Saxena, K.~Lerman, and A.~Galstyan.
\newblock A survey on bias and fairness in machine learning.
\newblock {\em ACM Computing Surveys (CSUR)}, 54(6):1--35, 2021.

\bibitem{northpoint2012}
Northpoint.
\newblock Compas risk \& need assessment system: Selected questions posed by
  inquiring agencies., 2012.

\bibitem{pcra2018}
A.~O. of~the United States Courts~Probation and P.~S. Office.
\newblock An overview of the federal post conviction risk assessment, 2018.

\bibitem{patel2008investigating}
K.~Patel, J.~Fogarty, J.~A. Landay, and B.~Harrison.
\newblock Investigating statistical machine learning as a tool for software
  development.
\newblock In {\em Proceedings of the SIGCHI Conference on Human Factors in
  Computing Systems}, pp. 667--676, 2008.

\bibitem{perkelReactiveReproducibleCollaborative2021}
J.~M. Perkel.
\newblock Reactive, reproducible, collaborative: Computational notebooks
  evolve.
\newblock {\em Nature}, 593(7857), May 2021.

\bibitem{polyzotis2019}
N.~Polyzotis, S.~Whang, T.~K. Kraska, and Y.~Chung.
\newblock Slice finder: Automated data slicing for model validation.
\newblock In {\em Proceedings of the IEEE Int' Conf. on Data Engineering
  (ICDE), 2019}, 2019.

\bibitem{rudinInterpretableMachineLearning2022}
C.~Rudin, C.~Chen, Z.~Chen, H.~Huang, L.~Semenova, and C.~Zhong.
\newblock Interpretable machine learning: {{Fundamental}} principles and 10
  grand challenges.
\newblock {\em Statistics Surveys}, 16, Jan. 2022.

\bibitem{dblpImpossibilityTheorem2020}
K.~K. S.
\newblock The impossibility theorem of machine fairness - {A} causal
  perspective.
\newblock {\em CoRR}, abs/2007.06024, 2020.

\bibitem{sagadeeva2021}
S.~Sagadeeva and M.~Boehm.
\newblock {\em SliceLine: Fast, Linear-Algebra-Based Slice Finding for ML Model
  Debugging}, p. 2290–2299.
\newblock Association for Computing Machinery, New York, NY, USA, 2021.

\bibitem{saha2020measuring}
D.~Saha, C.~Schumann, D.~Mcelfresh, J.~Dickerson, M.~Mazurek, and M.~Tschantz.
\newblock Measuring non-expert comprehension of machine learning fairness
  metrics.
\newblock In {\em International Conference on Machine Learning}, pp.
  8377--8387. PMLR, 2020.

\bibitem{skeem2016risk}
J.~L. Skeem and C.~T. Lowenkamp.
\newblock Risk, race, and recidivism: Predictive bias and disparate impact.
\newblock {\em Criminology}, 54(4):680--712, 2016.

\bibitem{verma2018fairness}
S.~Verma and J.~Rubin.
\newblock Fairness definitions explained.
\newblock In {\em 2018 ieee/acm international workshop on software fairness
  (fairware)}, pp. 1--7. IEEE, 2018.

\bibitem{wexler2020googleWhatIf}
J.~Wexler, M.~Pushkarna, T.~Bolukbasi, M.~Wattenberg, F.~Viégas, and
  J.~Wilson.
\newblock The what-if tool: Interactive probing of machine learning models.
\newblock {\em IEEE Transactions on Visualization and Computer Graphics},
  26(1):56--65, 2020. doi: {{%
10\hspace{.1pt}\discretionary{.}{%
}{.}\hspace{.4pt}1109\discretionary{/}{%
}{/}TVCG\hspace{.1pt}\discretionary{.}{%
}{.}\hspace{.4pt}2019\hspace{.1pt}\discretionary{.}{%
}{.}\hspace{.4pt}2934619}}


\bibitem{zhe2021protected}
Z.~Yu.
\newblock Fair balance: Mitigating machine learning bias against multiple
  protected attributes with data balancing.
\newblock {\em CoRR}, abs/2107.08310, 2021.

\bibitem{zemel2013learningFair}
R.~Zemel, Y.~Wu, K.~Swersky, T.~Pitassi, and C.~Dwork.
\newblock Learning fair representations.
\newblock In {\em International conference on machine learning}, pp. 325--333.
  PMLR, 2013.

\bibitem{zhang2018manifold}
J.~Zhang, Y.~Wang, P.~Molino, L.~Li, and D.~S. Ebert.
\newblock Manifold: A model-agnostic framework for interpretation and diagnosis
  of machine learning models.
\newblock {\em IEEE transactions on visualization and computer graphics},
  25(1):364--373, 2018.

\end{thebibliography}
\end{document}